\documentclass[conference]{IEEEtran}
\IEEEoverridecommandlockouts
\usepackage{cite}
\usepackage{amsmath,amssymb,amsfonts}
\usepackage{algorithmic}
\usepackage{graphicx}
\usepackage{textcomp}
\usepackage{xcolor}
\usepackage{multirow}
\usepackage{tabularx}
\newcolumntype{g}{X}                                \newcolumntype{s}{>{\hsize=.5\hsize}X}              \newcolumntype{Y}{>{\centering\arraybackslash}X}    
\usepackage{multicol}

\usepackage{hyperref}
\usepackage[aboveskip=4pt, belowskip=4pt]{subcaption}

\graphicspath{{figs/}{figures/}{pictures/}{images/}{./}} 

\usepackage{tabu}                      \usepackage{booktabs}                  \usepackage{lipsum}                    \usepackage{mwe}                       

\usepackage{mathptmx}

\begin{document}
	\title{A Tool for the Procedural Generation of Shaders\\using Interactive Evolutionary Algorithms
\thanks{Contact authors: Elio Sasso (elio.sasso@mail.polimi.it); Daniele Loiacono (daniele.loiacono@polimi.it; 0000-0003-4579-2606), Pier Luca Lanzi (pierluca.lanzi@polimi.it; 0000-0002-1933-7717).}
	}
	\author{\IEEEauthorblockN{Elio Sasso,  Daniele Loiacono,  Pier Luca Lanzi}\\		
		\IEEEauthorblockA{Dipartimento di Elettronica, Informazione e Bioingegneria (DEIB) - Politecnico di Milano, Milan, Italy}
	}

\maketitle
\begin{abstract}
We present a tool for exploring the design space of shaders using an interactive evolutionary algorithm integrated with the Unity editor, a well-known commercial tool for video game development. Our framework leverages the underlying graph-based representation of recent shader editors and interactive evolution to allow designers to explore several visual options starting from an existing shader. Our framework encodes the graph representation of a current shader as a chromosome used to seed the evolution of a shader population. It applies graph-based recombination and mutation with a set of heuristics to create feasible shaders. The framework is an extension of the Unity editor; thus, designers with little knowledge of evolutionary computation (and shader programming) can interact with the underlying evolutionary engine using the same visual interface used for working on game scenes. \end{abstract}


\section{Introduction}

Technical artists are highly specialized professionals with an in-depth knowledge of the graphic pipeline, advanced programming skills, and artistic sensibility; such professionals are known as shader artists. 
Small development teams usually do not have sufficient resources to hire such specialized profiles, and shader artist positions are often outsourced. 
Several visual tools have been developed to bridge the gap between artists and shader programming, such as 
	Shader Forge \cite{shaderforge}
Amplify Shader Editor \cite{amplifyshader},
Unity’s Shader Graph \cite{UnityShaderGraph},
and Unreal Engine Material Editor \cite{UnrealEngineMaterialEditor}.
These tools provide artists with abstraction over the underlying render pipeline 
	and replace the complex C/C++ syntax with visual editors.

In this paper, we present a tool for exploring the design space of shaders that combines an established visual tool for shader programming, Unity’s Shader Graph, and an interactive evolutionary algorithm into an extension of the Unity editor.\footnote{\url{http://www.unity.com}}
The tool starts from an existing shader in a Unity project, represented in the Unity Shader Graph format; it encodes it as a chromosome used to seed an initial population of shaders. Artists can explore the current shader population and evaluate individuals to provide qualitative feedback for the generation of the next population; graph-based mutation and recombination are integrated with heuristics to guide the generation of feasible (valid) shaders. Since our tool is an extension of the Unity editor, the evolutionary process is transparent to the users, who can simply browse a list of suggested shaders, submit their qualitative evaluation in terms of thumbs up, thumbs down, and save the shaders they prefer.

\section{Shader Programming and Visual Editors}
\label{sec:shader_programming}

In recent years, there has been a significant effort to integrate shader visual editors into popular video game development frameworks. Shader Weaver \cite{shaderweaver} is a third-party extension for Unity, which implements a visual editor for shaders for 2D rendering. Shader Forge \cite{shaderforge} was a visual shader editor also available as an extension of the Unity editor (supported until 2018), that provided a real-time preview of the edited shaders. It implemented helper nodes that would add complex shader features such as screen-space refraction and the possibility of user-defined per-light custom function, enabling more technical users to implement their custom lighting processes. The tool mainly focused on simplifying the shader creation process and streamlining more complex shader functions such as edge detection, vertex animation, and transparency effects. Once the user completed the shader creation, the tool would generate an HSLS file, which Unity could parse as a custom shader.  Amplify Shader Editor (ASE) is another visual editor for shaders available as an extension for Unity that is still supported and updated. It supports the latest render pipelines and provides greater control over shader functionalities \cite{amplifyshader}. Like Shader Forge, it also uses a graph-based representation of shaders. It offers more features when it comes to shader editing than Shader Forge and the visual editor for shaders currently integrated into Unity.
 
\section{Related Work}
\label{sec:related}
Few works have applied evolutionary computation to the generation of shader code \cite{DBLP:conf/eurogp/EbnerRA05,pcgvertex,pcgfragment,10.1145/2070781.2024186,nips2018:cppns}. All these works 
focus the search on a single specific single aspect of the shader (in contrast our framework consider the entire pipeline) and 
are based on GLSL, whereas our framework uses the abstract graph-based representation used by the most advanced shader editors \cite{shaderforge,shaderweaver,amplifyshader}. Most of them use tree structures to represent shaders (our framework also uses a tree-based representation).

Ebner et al. \cite{DBLP:conf/eurogp/EbnerRA05} applied genetic programming to evolve vertex and pixel shaders via user interaction. Shaders are encoded as linear sequences of commands which are translated into a high level computer graphics (CG) language. Similarly to what we do in our system, individuals are applied to four different objects that are then presented to users who can select which individuals should reproduce. Note that, the representation is based on a custom representaiton of commands that must be then translated into actual shader. Accordingly, users can examine (and possibly modify) only the phenotype not the genotype that has produced the shader. In contrast, our system work on the native shader representation so that in any moment users can access the shader population and (if they want) they can modify any individual. 

Quiroz and Dascalu \cite{pcgvertex} applied genetic algorithms to the code of a vertex shader to explore the space of different types of vertex displacement that can be applied to the rendered mesh. Vertex displacement modifies the visual aspect of a mesh without working on its actual geometry. For example, it can distort the aspect of a mesh without working on its underlying representation.
Accordingly, the evolution in \cite{pcgvertex} tried to modify the visual aspect of an existing mesh by applying an offset to each vertex of the original mesh. They developed a web application to render a mesh using a variety of shaders generated through evolutionary algorithms. The user could select the individual in the population that best fit their needs as the seed for the next generation. Quiroz and Dascalu \cite{pcgvertex} considered only one specific aspect of shader code (vertex displacement shaders), and many of the iterations produced destructive modifications of the mesh. Furthermore, the web application could only work on the entire mesh and was not able to focus on specific characteristics. In contrast, our tool works on any component of the shader encoded in the graph-based representation and, by acting on single nodes, can work on specific (local) aspects of the shader. 

Vertex shaders can manipulate shapes, visuals, and some lighting characteristics of a rendered object; instead, fragment shaders (also called pixel shaders) are responsible for the final color of the object, shading included, which usually makes up most of the appearance of the rendered mesh. 
Howlett, Colton, and Browne \cite{pcgfragment} applied an interactive genetic algorithm to fragment shaders to render the landscapes generated in the game Subversion \cite{subversion} with different shades of color, which would highlight some areas rather than others. In Subversion \cite{subversion}, players are presented with an aerial view of a procedurally generated city which, by default, is rendered in black and white black with white edges.
The genetic algorithm in \cite{ pcgfragment} focused on changing the shades of color of different parts of the city.
The fitness function was based on the hue, saturation, and brightness of the color palette in the rendered image. Evolution could evolve different parameters for different areas of the city. A carefully chosen heuristic would set the final shader to explore the search space in a way optimized for the final use case. Also in this case, the search space was restricted to a particular shader feature. In contrast, our tool can work on any aspect of the shader available in the existing graph-based representation.

Previous approaches developed shaders that could work on single elements of the rendered scenes (e.g., materials, meshes). In contrast, \cite{DBLP:journals/cgf/NalbachAMSR17,nips2018:cppns} focused on shaders that work on the entire screen and thus reproduce effects over the entire rendered scene. Nabach et al. \cite{DBLP:journals/cgf/NalbachAMSR17} applied convolutional neural networks to learn screen space effects (that is, screen space shading) from examples without the need of human programming. Snelgrove and Tesfaldet \cite{nips2018:cppns} also focused on screen space effect and applied Compositional Pattern Producing Networks (CPPNs) to create effects over the entire rendered image. 

Finally, Sitthi-Amorn et al. \cite{10.1145/2070781.2024186} applied genetic programming to simplify procedural shaders to expose the inherent trade-off between speed and accuracy. They compared their approach with existing methods for pixel shader simplification \cite{10.5555/844174.844176, 10.1145/1073204.1073212} showing that their system worked on a wider space of code transformations and produced faster and more accurate results. Note that,  \cite{10.1145/2070781.2024186} focuses on simplifying an existing shader aiming at a balance between performance and visual accuracy instead of generating new shaders like \cite{pcgfragment,pcgvertex} and our tool. 
 
\section{Unity ShaderGraph}
\label{sec:shadergraph}
Unity is a game engine released in June 2005 by Unity Technology that, over the years, has become one the go-to tool for cross-platform development of video games and other multimedia applications. It has a modular architecture that has been steadily extended with several tools that added advanced functionalities to the platform. In recent years, high-end visuals have been a major focus of engine development with the introduction of the scriptable render pipeline, the high definition render pipeline (HDRP), the universal render pipeline (URP), and more recently ShaderGraph, a visual editor for shaders conceptually similar to the Amplify Shader Editor \cite{amplifyshader} and Shader Forge \cite{shaderforge}.

\begin{figure}[t]
	\includegraphics[width=\columnwidth]{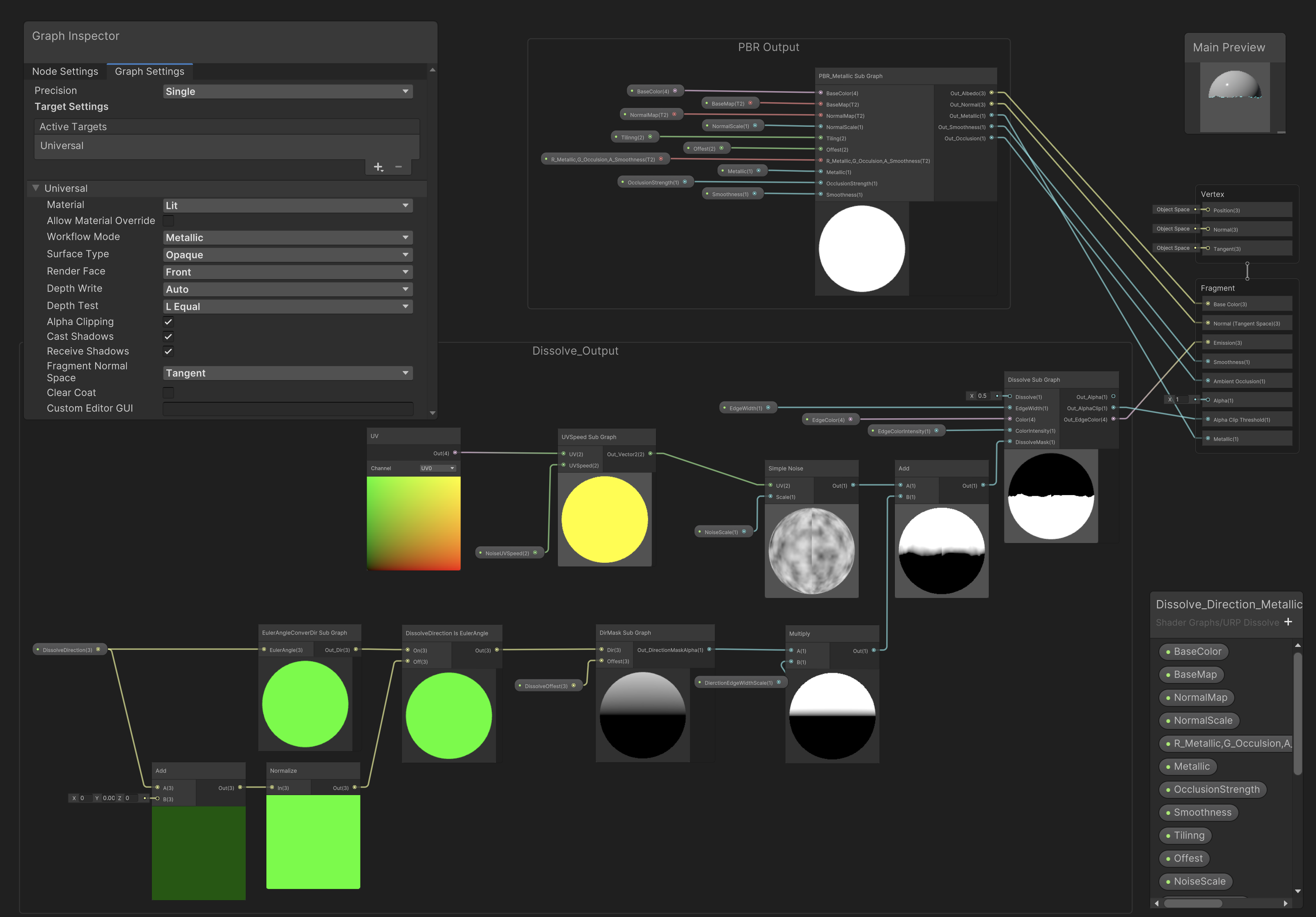}
	\caption{Unity Shadergraph graph inspector view of dissolve direction metallic shader for Unity Universal Render Pipeline \cite{dissolve_shader}. }
	\label{fig:shadergraph}
\end{figure}

ShaderGraph is a visual programming tool for writing shaders for the Unity render pipelines. 
It represents complex shaders as graphs modeling the flow of information inside the render pipeline (Figure~\ref{fig:shadergraph}). Nodes represent elementary operations in the different stages of a shader and encode one or more lines of code in the final shader run by the GPU.
Users can add and remove nodes, modify their parameters (e.g., change precision of the float values to optimize the shader performance), as well as adding new input values to the shader itself. 
The Graph Inspector panel (Figure~\ref{fig:shadergraph} top left corner) presents users with a detailed list of node settings and let users modify the shader's mode of operation (e.g., the target render pipeline or the precision for all the nodes in the graph). 
Shader inputs can be accessed from an external source (e.g., in-game code) and changed at run time without modifying other section of the renderer. 
At the end, the graph is translated into shader code that can be used by the Unity engine render pipelines (Figure~\ref{fig:shadergraph_code}).

\subsection{Access to the Shader Pipeline}
The graph representation abstracts the underlying (complex) render pipeline and exposes only relevant sections that users can modify like for example, 
(i) the vertex position, modifying the 3D coordinates of any mesh vertex;
(ii) the normal vector (at per-vertex basis), responsible for computing lighting data such as reflection and shadows at run time;
(iii) the tangent vector, usually modified along with the normal vector to maintain consistency; 
(iv) the base (albedo) color that dictates either the final color of the fragment or the color before operating light computations;\footnote{This is the parameter which is changed in most of the use cases, as it is often used with texture maps which gets sampled on a per-pixel level.}
(v) the normal (at fragment stage) used to the recompute the normal vector using interpolated per-pixel values, for example when  shadow details are reintroduced into the final image through the use of normal map sampling \cite{10.1145/800248.507101};
(vi) smoothness and metallic, used to emulate real-life objects with a high degree of detail when applying physical base rendering to meshes;
(vii) ambient occlusion, applied for reproducing shadows for small nooks in meshes that would be too expensive to compute; thus, ambient occlusion can darken the final color by a preferred amount and, similarly to normal mapping, occlusion maps are used in order to introduce an extra level of detail;
(viii) emission color, that takes maps as input in order to show lit up sections of the rendered mesh;
(ix) alpha and alpha clip threshold values used in all shaders to render transparent and translucent objects. 
All of these values contribute to the final computation of pixel colors produced by the rendering process  and their effect can change based on the selected shader type.
Shader Graph also let users define custom nodes implemented using traditional shader programming.

\subsection{Shader Code Generation}
\begin{figure*}
	\centering
\includegraphics[width=.9\textwidth]{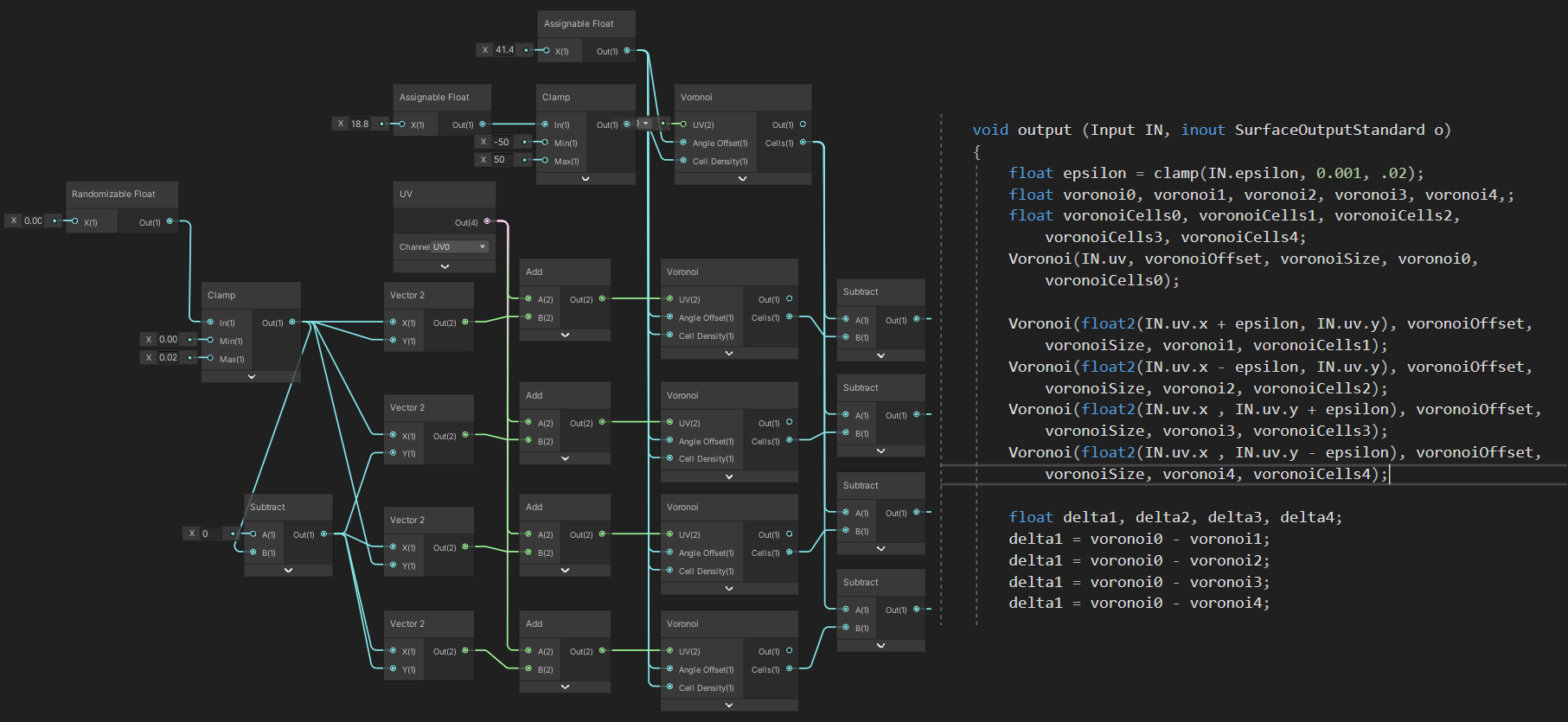}
		\caption{Graph representation of a section of shader (left) with the same shader coded manually. As can be noted the graph representation is bloated with respect to the actual code needed to implement the same visual effect.}
		\label{fig:shadergraph_code}
\end{figure*}

Shader Graph parses the graph and generates code snippets based on the graph nodes. For each node, all the inputs are mapped to variables in the shader code and fed into pre-built functions defined for each node representation. 
The same process is applied to outputs which are also mapped to internal variables. 
The generation process creates a fully functioning shader with a slight performance overhead due to abundance of extra variable declarations and the presence of definition of functions which only wrap HLSL functions. Moreover, all function calls and operations in the shader's code are represented by single separate nodes, which can bring the resulting graph to bloat quite rapidly compared to a manually written code operating in the same way (Figure \ref{fig:shadergraph_code}).
Accordingly, the shader code generated by Shader Graph include code fragments that are not represented in the graph but are needed to ensure the highest possible degree of flexibility without encumbering the users with technical parameters that contribute little to the end product.

\section{Interactive Shader Evolution}
\label{sec:interactive}
Visual tools for shader programming have helped bridge the gap between the design of custom and dedicated shaders among less technical users by providing an abstraction over the complexity of the render pipeline. Nevertheless, graph-based representation still requires technical knowledge about the rendering workflow and can rapidly scale up in size and complexity. Therefore, artists might find it challenging to explore the wide variety of options the tool offers. Our goal was to provide artists and less technical users with a way to explore the shaders' design space and iterate over attractive visual options by introducing a layer of abstraction on top of the shader creation process. 

\subsection{Unity Graph-based Representation}
Unity does not provide APIs to access ShaderGraph data nor its underlying representation. Accordingly, we initially studied Unity Shader Graph format and developed a library to access all the functionalities available in the shader editor, such as (i) creating and deleting all the node types available in ShaderGraph; (ii) representing the feasible nodes and shader inputs and outputs; (iii) connecting nodes with several kinds of edges; (iv) describing materials, etc. 
In addition, the library implements all the functions needed to manage shader populations and to implement the genetic programming operators including (i) the generation of feasible random graphs; (ii) the recombination of two shaders by swapping sections of their graph  representations; (iii) to analyze graph topologies (e.g., graph traversal to search for nodes that can be directly or indirectly connected to a target node); and (iv) to manage sets (populations) of shader graph files.

\subsection{Individual Representation}
\label{ssec:individual_representation}
Our evolutionary algorithm works on the native graph representation of shaders. In Unity's Shader Graph, a shader is a forest of small interconnected subtree structures, each one representing a section of the overall rendering process. Our interactive evolutionary algorithm uses the same representation with chromosomes encoding a forest of subtrees. Nodes are either connected to a shader input (that users can change from code), to another a shader node, or have their input slots associated with values that the users can specify or randomly generate from specialized noise generator nodes (similarly to ephemeral constants in genetic programming \cite{koza:1992}).

\subsection{Evolutionary Operators: Mutation and Crossover}
\label{ssec:operators}
The mutation operator is implemented using a scaffolding approach. It applies a set of predefined functions that modify the graph to ensure that the resulting graph still represents a valid shader. We implemented simple mutation functions that can randomly change the presets of nodes and expand existing subtrees. We also implemented mutation functions that apply domain knowledge specific to shaders. For example, we have a mutation function specific to noise functions (named Swap Noise Map). Shaders often include noise generator nodes (e.g., Voronoi, Simplex or Gradient Noise) to generate variations in the rendered mesh while avoiding repetitive patterns. When a shader graph has one or more noise functions, our mutation function can modify the noise function in use. A shader will sometimes contain different calls to the same noise function, all semantically bound to each other. 
Mutation keeps this caveat into account by examining the graph topology to determine what noise functions are semantically linked and swapping all such related noise nodes. 

Shaders are forests of many (typically small) interconnected trees, each one representing a functionality of the overall shader (Section \ref{ssec:individual_representation}). Crossover works on the connection among these small trees and applies crossover on the output nodes of two-parent trees. 
Thus, the crossover does not work on the structure of the single small subtrees in the forest that represents a single shader. Instead, it selects the output nodes in two subtrees and recombines the output nodes, thus connecting a section of the first parent shader to a section of the second shader and vice versa. Offsprings consist of two new graph topologies with subgraphs partly from one parent and partly from the other. For each input node in the resulting graph, the range of values the input node can take is inherited by the corresponding parent. Overall, crossover exchanges two portions of parent shaders, similar to what happens in tree-based genetic programming \cite{koza:1992} by working on the connections between subtrees.

\subsection{The Interactive Evolutionary Algorithm}
We implemented a steady-state interactive evolutionary algorithm. Users can start from an initial population seeded from a set of existing shaders or a completely random population. When generating a new shader graph, the procedure first determines whether the shader will be lit (it will process light information) or unlit (it will not process light information). 
Then, similarly to what is done in genetic programming, it applies mutations to generate a set of random shader trees and creates the shader input values. 
At each iteration, the initial population is presented to the users by showing the effect of each shader has on a default scene (similarly to what is done in  \cite{DBLP:conf/eurogp/EbnerRA05}). Users can score the shaders they like most in the population. Next, they can either select two shaders they wish to recombine and mutate or they can let the evolutionary algorithm select the two parents using tournament selection. The former option provides users with more agency over the process whereas the latter one promotes exploration of the design space. 
The selected shaders are recombined and mutated to generate two new offspring that are inserted in the population. Finally, two shaders are deleted to keep the population size constant.

\section{Integration in the Unity Editor}
\label{sec:the_tool}
The interactive evolutionary algorithm has been developed as an extension of the Unity Editor using the interface that Unity provides to add functionalities to their system.\footnote{\url{https://docs.unity3d.com/Manual/ExtendingTheEditor.html}} Thus, designers can edit their own shaders using Unity's editor, then save them and apply the evolutionary algorithm to explore interesting variations of their shaders. 

\subsection{Startup}
\begin{figure}
		\centering
		\includegraphics[width=0.8\columnwidth]{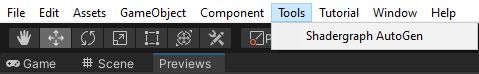}

		\vskip .1cm
		(a)

		\vskip .5cm
		\includegraphics[width=0.8\columnwidth]{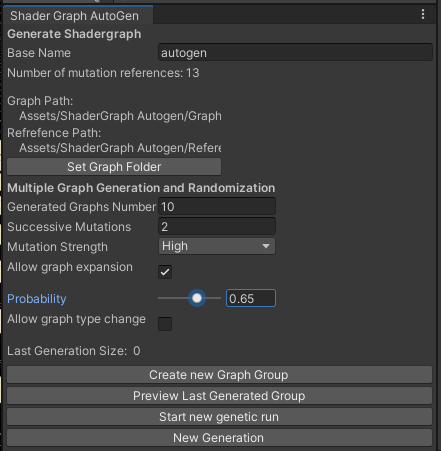}
		
		\vskip .1cm
	    (b)
		
	   \caption{Our tool can be launched by selecting the tools tab into Unity (a) 
	   	that opens the tool launch tab presents users with all the settings that users can modify (b).}
		\label{fig:toolwindow}
		\label{fig:tooltab}
		
\end{figure}

Our tool can be imported into a project using the standard Unity package manager. Once installed, users can launch it through a dedicated menu of the editor window (Figure \ref{fig:toolwindow}a). The tool opens a Unity tab inside the editor (Figure~\ref{fig:toolwindow}b) 
that let the user specify several pararameters including, (i) the path to the directory where new generated shaders will be saved; (ii) the number of graphs generated in each new generation (with a default value of two), that influences the amount of variety introduced in a single iteration of the algorithm; (iii) the number of mutation types that may be applied to an offspring after this has been selected for recombination; (iv) the mutation strength (with values low, medium, and high), that determines how much each mutation will impact the offspring shader. For example, a high strength mutation will modify parts of the graph flowing into the most crucial output nodes such as the Base Color (or Albedo) of the shader and the computed normal vectors, which sharply affect reaction to light by the rendered mesh. Users can modify the mutation strength at each generation, allowing them to progressively focus evolution toward certain types of shaders; 
(v) the graph expansion toggle (and the corresponding probability), that can disable the possibility to expand graphs in the population by adding trees; this option was introduced as another way to focus evolution on specific individuals in the population since, when disabled, it will focus mutation on the shader parameters not on its structure. The panel also contains buttons that let the user launch a new run, generate new offspring from the selected graphs and preview the last generated population. 

\subsection{Interaction}
\begin{figure}
	\centering
	\begin{tabular}{cc}
		\parbox{.5\columnwidth}{
			\includegraphics[width=.45\columnwidth]{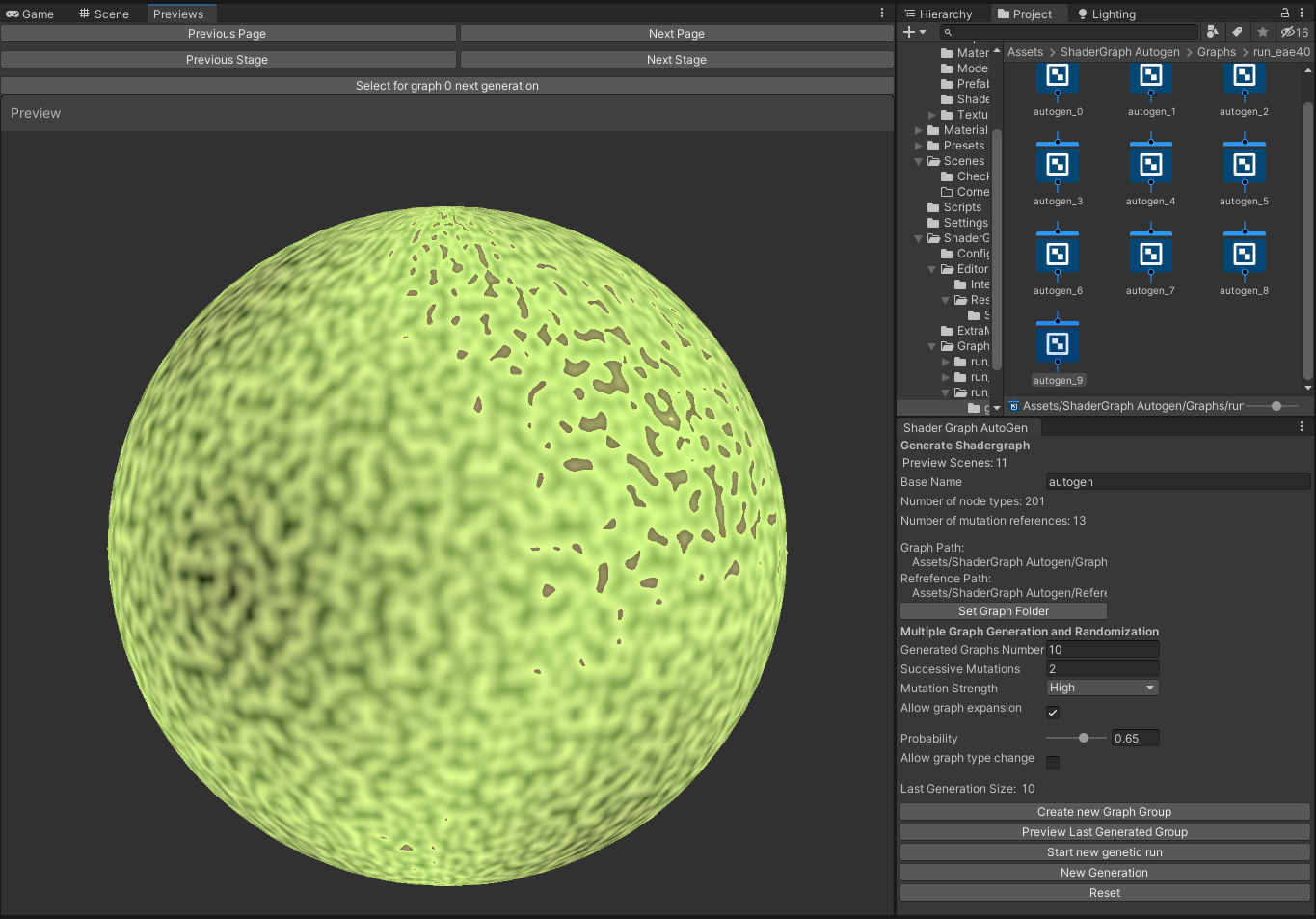}
		}
	 & 
		\parbox{.5\columnwidth}{
			\includegraphics[width=.45\columnwidth]{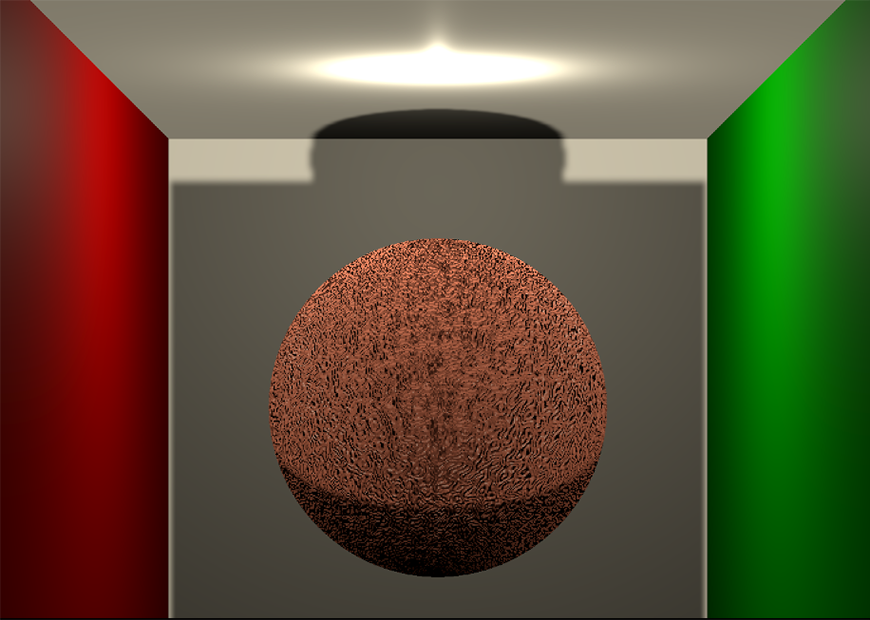}
		} \\
	(a) & (b) \\
	&  \\
	
		\parbox{.5\columnwidth}{
			\includegraphics[width=.45\columnwidth]{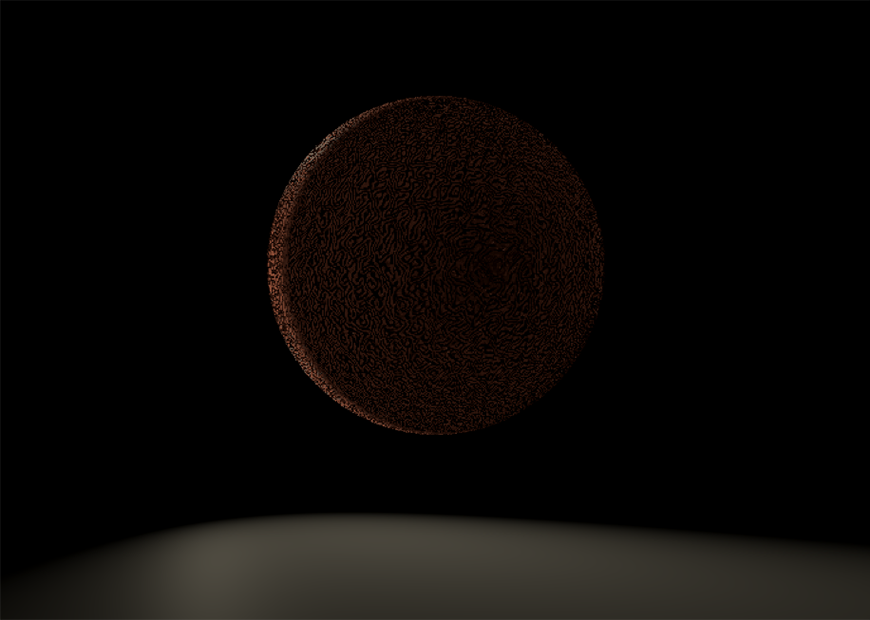}
		} 
		
		& 
		\parbox{.5\columnwidth}{
			\includegraphics[width=.45\columnwidth]{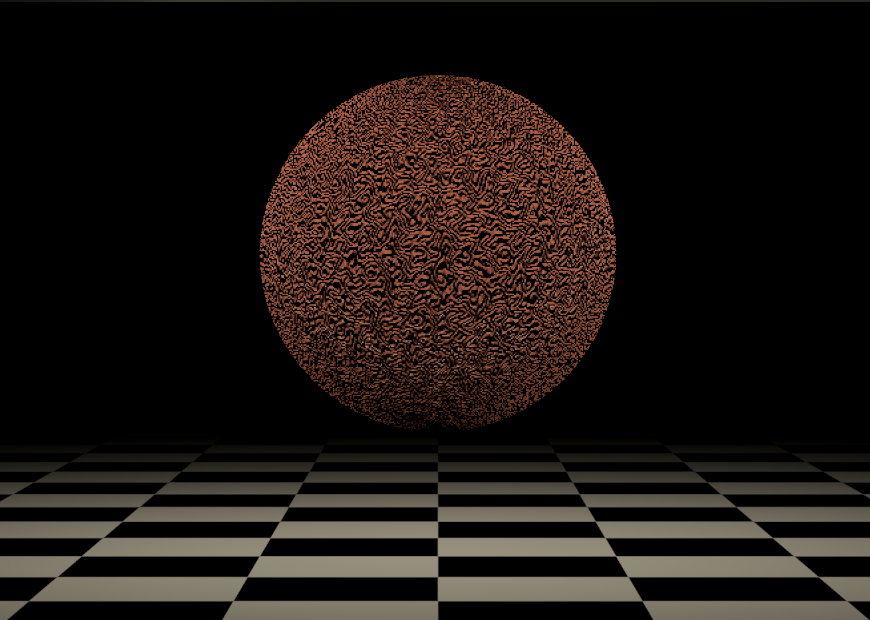}
		} \\
	(c) & (d) \\
	& \\

	\end{tabular}	
	
	\caption{Shader preview screen using (a) Unity standard shader view; 
		(b) Cornell Box; 
		(c) darkroom with a moving light;
		(d) traditional checkerboard floor.}
	
	\label{fig:previews}
\end{figure}

At the start of a run, the initial population is created based on the settings specified in the editor window. Users can explore all the generated shaders can be presented using a preview window showing how the shader renders a lighted sphere (Figure \ref{fig:previews}a), which is Unity's default visualization for shaders in the editor inspector. Users can select the number of previews appearing on the screen based on their screen size, the complexity of the shaders they are working with, and the available computer power. When the population size is larger than the number of previews on screen, the shaders are organized over more pages that the user can navigate. Users can interact with the shader previews and select another mesh for the rendering (e.g., a box, a capsule, or any custom mesh) using the \textit{Custom Mesh} menu when right-clicking on the preview window. This functionality is, for example, useful when working with shaders that use vertex displacement, which modifies the object appearance by shifting the position of the mesh vertices. Users can also preview shaders under different conditions. 

Our tool provides three typical preview scenarios that can be selected through the preview window: (i) the classic Cornell Box (Figure~\ref{fig:previews}b), \footnote{\url{https://en.wikipedia.org/wiki/Cornell_box}} which shows how the shader will react to fixed lighting and wall introducing different color reflections; (ii) a dark room with moving lights (Figure~\ref{fig:previews}c), which shows how the procedural normal vectors react to light; and (iii) the traditional checkerboard ground with fixed lighting (Figure~\ref{fig:previews}d).

Users can score the shaders they like most using the preview window to provide qualitative feedback to the underlying evolutionary engine. When all the shaders have been scored, tournament selection is applied to select two parent shaders that are copied, recombined, and mutated. Alternatively, users can directly select two shaders for recombination and mutation. We introduced this second option to give more agency to users interested in actively participating in the selection, recombination, and mutation procedures. The two offspring are inserted in the population and will appear on the screen (using the same preview window). Two shaders will be randomly selected from the population for deletion. Users can save the shaders they like most at any time, thus implementing elitism. 
All the individuals are saved using the standard Unity shader graph format into the dedicated directory specified at the beginning. Accordingly, at any time, users can open any shader in the population using Shader Graph; they can import it into an existing Unity project to test it thoroughly; they can also check the code generated by Unity.

\section{Conclusions}
\label{sec:conclusions}
We developed a tool for exploring the design space of shaders using an interactive evolutionary algorithm that we seemingly integrated into Unity---the well-known IDE for video game and multimedia application development. The tool combines an established visual tool for shader programming, Unity’s Shader Graph, and an interactive steady-state evolutionary algorithm into an extension of the Unity editor. It can help artists with limited knowledge of the render pipeline approach the shader creation and editing, starting from an existing shader (used to seed an initial population) or a completely random population. Shader programming, using code or visual tools, requires in-depth knowledge of the rendering process. Sometimes, it might be difficult for artists and designers to develop intriguing variations of existing visual effects. Our tools help users explore the space of visual possibilities by letting them provide qualitative feedback about the visuals they prefer to the underlying evolutionary engine. Such feedback is then used to select, recombine, and mutate shaders in the population to create new visual effects. Users can also take an active role in the process and ask the engine to recombine and mutate shaders that they explicitly selected. This option gives more agency to the users, which was very appreciated in a preliminary evaluation we performed with few human subjects. The evolutionary algorithm works on the native graph-based representation used by Unity Shader Graphs, which represents shaders using forests of interconnected small trees. Recombination works by modifying the connection between existing subtrees. The mutation is implemented, similarly to what is done in genetic programming, using a set of predefined functions that can modify any subtree in the forest. 

\vfill\eject
We performed a preliminary evaluation with a limited number of human subjects, mainly design and engineering students enrolled in video game design courses. The feedback is promising, but evaluation with a larger population is needed. Unity does not provide a native interface to access shader graph representation. Thus, our library will have to be updated if future releases of Unity Shader Graph. Accordingly, we plan to open-source the project to be able to get help to maintain the tool with future Unity releases.
 

\end{document}